\pdfoutput=1
\documentclass[%
 reprint,
 superscriptaddress,
 amsmath,amssymb,
 aps,
floatfix,
]{revtex4-1}

\usepackage{graphicx}
\usepackage{dcolumn}
\usepackage{color}	
\usepackage{bm}
\usepackage{braket}

\begin{document}


\title{An \textit{ab initio} study of shock-compressed copper}

\author{Maximilian \surname{Schörner}}
\affiliation{%
 University of Rostock, Institute of Physics, 18051 Rostock, Germany
}%

\author{Bastian B. L. \surname{Witte}}

\affiliation{%
 University of Rostock, Institute of Physics, 18051 Rostock, Germany
}%

\author{Andrew D. \surname{Baczewski}}

\affiliation{%
 Center for Computing Research, Sandia National Laboratories, Albuquerque NM, USA
}%

\author{Attila \surname{Cangi}}

\affiliation{%
 Center for Advanced Systems Understanding (CASUS), Helmholtz-Zentrum Dresden-Rossendorf, D–02826 Görlitz, Germany
}%

\author{Ronald \surname{Redmer}}

\affiliation{%
 University of Rostock, Institute of Physics, 18051 Rostock, Germany
}%

\date{\today}

\begin{abstract}
We investigate shock-compressed copper in the warm dense matter regime by means of density functional theory molecular dynamics simulations. We use neural-network-driven interatomic potentials to increase the size of the simulation box and extract thermodynamic properties in the hydrodynamic limit. We show the agreement of our simulation results with experimental data for solid copper at ambient conditions and liquid copper near the melting point under ambient pressure. Furthermore, a thorough analysis of the dynamic ion-ion structure factor in shock-compressed copper is performed and the adiabatic speed of sound is extracted and compared with experimental data.
\end{abstract}

\pacs{Valid PACS appear here}
\maketitle


\section{Introduction}
\label{sec:intro}

Understanding matter in extreme conditions is challenging. Especially warm dense matter (WDM), which is characterized by temperatures above a few electronvolt (eV) and solid densities exhibits non-negligible degeneracy and strong correlations that must be treated in a quantum mechanical many-body framework~\cite{Graziani,Glenzer2009}.  Experimentally, due to their high energy density, these states can only be created transiently, and therefore, must be probed on short time scales using intense short-wavelength radiation. Shock-compression experiments are among the premier ways extreme conditions can be reached in the laboratory. They have been used to study the high-pressure phase diagram of various geological materials~\cite{Duffy2019}, metals like silver, gold and platinum~\cite{Briggs2019,Sharma2019,Sharma2020}, iron at super-Earth conditions~\cite{Kraus2022}, and hydrocarbons~\cite{Kraus2018,Hartley2019,Luetgert2021}, even revealing novel phenomena like the formation of diamonds in the interior of Neptune~\cite{Kraus2017}.
In shock and ramp compression studies, copper itself is often used as a resistivity gauge~\cite{Golyshev2011}, and its behavior under extreme conditions has been the target of several theoretical and experimental studies over the past decades. The conductivity of expanded liquid copper has been studied in rapid wire evaporation experiments~\cite{DeSilva1998} and isochoric heating experiments using a closed vessel apparatus~\cite{Clerouin2012}, while the effects of femtosecond irradiation has been studied using first-principles calculations~\cite{Smirnov2020}. Melting curves over a wide pressure range have been predicted theoretically and measured~\cite{Errandonea2013,Hayes2000,Wu2011,Moriarty1987,Belonoshko2000}. Recently, Baty~\textit{et~al.}~\cite{Baty2021} have used \textit{ab~initio} simulations to study the melting line of copper up to pressures relevant for shock compression, while accounting for the experimentally and theoretically predicted meta-stable bcc phase~\cite{Bolesta2017,Sims2019,Sharma2020_2,Smirnov2021}. Furthermore, a plethora of shock wave measurements in the Mbar regime are available, although the uncertainties for measurements beyond 5~Mbar increase significantly. Notoriously, it is challenging to extract reliable information on structural or transport properties at these extreme conditions. However, with novel improvements to the spectral resolution at high-brilliance x-ray free electron laser facilities, it is now becoming possible to measure the ion dynamics of transient WDM states through inelastic x-ray scattering~\cite{McBride2018,Descamps2020,Wollenweber2021}, as well as structural changes via x-ray diffraction~\cite{Kraus2018,Hartley2019,Luetgert2021}. While a lot of theoretical work regarding copper has been performed on phase transitions and the energy transfer from the electrons to ions, we focus, here, on the ion dynamics that can be accessed in scattering experiments.

The dynamic structure factor (DSF) is vital for accurately describing the dynamics of matter under extreme conditions. There has been a lot of work on the DSF in the context of molecular dynamics (MD) simulations, ranging from the direct observation from ion trajectories~\cite{Allen1989,Alemany1999} to fits to analytical expressions~\cite{Mryglod1998,Bryk2011,Hansen2006}. Large-scale classical simulations have been employed to reach the hydrodynamic regime~\cite{Mithen2011,Mithen2014,Cheng2020} and to study structural properties and propagation in glasses and disordered solids via the disperison of the longitudinal and transverse DSF~\cite{Monaco2009,Mizuno2014,Oyama2021}. Furthermore, MD simulations have been coupled to density functional theory~\cite{Hohenberg1964,Kohn1965} (DFT-MD) to reach \textit{ab~initio} accuracy in the description of the DSF~\cite{Alemany2004}. This allows us to probe the ion dynamics of dynamically compressed targets, requiring sophisticated many-body simulations, that take into account the quantum mechanical nature of WDM states, to compare with the experimental observations. These DFT-MD simulations have proven successful at describing the principal shock Hugoniot~\cite{Knudson2015,Millot2018,Ravasio2021} and the ion-ion DSF of various materials~\cite{Witte2017, Rueter2014, White2013, Gill2015}. Recently, the use of neural network potentials has emerged, combining the benefits of the large-scale classical simulations with the \textit{ab~initio} accuracy of the forces in DFT. Several studies have shown the application of this technique to the DSF~\cite{Cheng2020,Zeng2021,Schoerner2022} and studied the extent of the hydrodynamic regime and the accessibility of various transport and thermodynamic properties.

Here, for the first time, we apply these state-of-the-art methods to shock-compressed copper at experimentally reachable conditions and we make predictions for experimentally observable quantities like the static and dynamic ion-ion structure factor. First, we benchmark our results against experimental results for solid and liquid copper, and then compute the principal Hugoniot curve and study the change of the ion dynamics. A brief summary of the relevant equations and the details of the simulation methods are given in Sec.~\ref{sec:theory}. In Sec.~\ref{sec:solid-metal}, we determine the phonon spectrum of solid copper at ambient conditions from the ion-ion DSF and compute the dynamic electrical conductivity. For liquid copper at ambient pressure near the melting point, we compute the static and dynamic structure factor and compare to experimental results, see Sec.~\ref{sec:liquid-metal}. Subsequently, we compute the principal Hugoniot curve in Sec.~\ref{sec:hugo} and compare to shock compression experiments, and experimental and theoretical predictions for the melting line. Finally in Sec.~\ref{sec:shock-copper}, we study the evolution of the static and dynamic ion-ion structure factor during shock compression and extract the adiabatic speed of sound, which we compare to recent measurements by McCoy~\textit{et~al.}~\cite{McCoy2017}.

\section{Theoretical method}
\label{sec:theory}

Through DFT-MD simulations, we gain access to the time-dependent ion positions $\vec{r}_i(t)$ and velocities $\vec{v}_i(t)$.
By virtue of the Wiener-Khinchin theorem~\cite{Wiener, Khinchin}, the dynamic ion-ion structure factor
\begin{equation}\label{eq_Sii_WK}
S_{\mathrm{ii}} ( \vec{k},\, \omega) = \frac{1}{2\pi N} \int_{-\infty}^{\infty} \mathrm{d}t \, \langle n_{\vec{k}}(\tau) \, n_{-\vec{k}} (\tau + t)  \rangle_{\tau} \, e^{i\omega t},
\end{equation}
which contains all information on the dynamics of the ion system, can be defined. Here $N$ is the number of ions in the system, $\vec{k}$ is the wave vector, $\omega$ is the frequency, and the spatial Fourier component of the ion density $n(\vec{r}, t)$ is given as
\begin{equation}\label{eq_ion_dens}
n_{\vec{k}} (t) = \int_{\mathbb{R}^3} \mathrm{d}^3r \, n(\vec{r}, \, t) \, e^{i\vec{k} \cdot \vec{r}} = \sum_{i=1}^{N} e^{i \vec{k} \cdot \vec{r}_i(t)} \, ,
\end{equation}
\begin{equation}
n (\vec{r}, t) = \sum_{i=1}^N \delta^3 (\vec{r} - \vec{r}_i (t)) \, ,
\end{equation}

with the time-dependent ion positions $\vec{r}_i(t)$. In Eq.~\eqref{eq_Sii_WK}, $\tau$ denotes an absolute time relative to the time delay $t$. The ensemble average, denoted by subscript $\tau$, is taken to be the sample average for independent configurations with different values of $\tau$ but the same value of $t$.

According to the hydrodynamic model~\cite{Hansen2006}, the dispersion of the collective side peak of $S_\mathrm{ii} ( \vec{k}, \omega )$, called sound mode in the hydrodynamic limit, is connected to the adiabatic speed of sound $c_\mathrm{s}$ via
\begin{equation}
 \omega_\mathrm{sound} (\vec{k}) = c_\mathrm{s} |\vec{k}| \, .
\end{equation}
The position of the sound mode $\omega_\mathrm{sound}$ can be extracted from the DSF via a fitting scheme (for details, see Refs.~\cite{Rueter2014,Witte2017,Schoerner2022}).

The dispersion of the longitudinal current-current correlation spectrum $J(\vec{k}, \omega)$, which is closely related to the DSF via 
\begin{equation}\label{eq_current_correl}
J(\vec{k}, \, \omega) = \frac{\omega^2}{k^2} S_\mathrm{ii} ( \vec{k}, \, \omega ),
\end{equation}
defines the apparent sound speed $c_\mathrm{l}$.

The long-wavelength limit of the static ion-ion structure factor, which is defined as the frequency integrated ion-ion DSF $S_{\mathrm{ii}} (k) = \int_{-\infty}^{\infty} S_{\mathrm{ii}} (k,\omega) \, d\omega$, can also be determined from
\begin{equation}\label{eq-Sk0}
\lim_{k \to 0} S_{\mathrm{ii}} (k) = \kappa_T n_\mathrm{i} k_{\mathrm{B}} T
\end{equation}
via thermodynamic relations~\cite{Chaturvedi}. Here, $n_\mathrm{i}$ is the ion density, $T$ is the temperature and $\kappa_T$ is the isothermal compressibility which is also accessible via the thermodynamic relation
\begin{equation}\label{eq_th_compress}
\kappa_T = - \frac{1}{V} \left( \frac{\partial V}{\partial P} \right)_T \ ,
\end{equation}
where $V$ is the volume of the simulation box and $P$ is the pressure.

In order to analyze shock compression experiments, we employ the Hugoniot equation~\cite{Hugoniot1887, Hugoniot1889, Rankine1851, Rankine1870}
\begin{align}\label{eq_hugoniot}
\epsilon_1 - \epsilon_0 = \frac{1}{2} & \left( P_1 +P_0 \right) \left( V_0 - V_1 \right), \\
\epsilon_a = & \frac{E_a}{m_a}, \quad a=0,1 \, ,
\end{align}
which can be derived from the conservation of energy~$E$, momentum~$p$, and mass~$m$ at a propagating shock front. Here, the subscript $0$ indicates the conditions of the unshocked material while subscript $1$ indicates the conditions of the shocked material.

The DFT-MD simulations in this work were performed within the Vienna Ab-Initio Simulation Package (VASP)~\cite{VASP1,VASP2,VASP3}. The electron density at each time step is computed according to the finite-temperature DFT approach~\cite{Mermin2}, using the generalized gradient approximation of Perdew, Burke and Ernzerhof~\cite{PBE} for the exchange correlation functional (XC-functional). The MD is carried out using the Born-Oppenheimer approximation by solving Newton's equations of motion for the ion positions. The forces are determined by the electronic charge density, where the electrons always remain in an instantaneous, thermal equilibration defined by the ion positions.

Within VASP the Kohn-Sham orbitals are expanded in a plane wave basis set up to a cutoff energy $E_\mathrm{cut}$, which we set at 800 eV. For the ion potential of copper a projector augmented-wave potential~\cite{Blochl1994} is used (PAW PBE Cu GW 19May2006), which treats the outer eleven electrons explicitly in the DFT framework, with the remaining electrons frozen in the core. For temperature control, the algorithm of Nos\'e{}-Hoover~\cite{Nose,Hoover} is used with a mass parameter corresponding to a temperature oscillation period of 40 time steps. 
To allow for melting and freezing along the principal Hugoniot curve, the simulation box for copper is spanned by lattice vectors of the face-centered cubic~(fcc) structure. For the simulations at ambient pressure, $125$ copper atoms were used, while for all conditions along the Hugoniot curve 64 copper atoms were placed in the simulation box.
The sampling of the Brillouin zone was carried out at the Baldereschi mean-value point~\cite{Baldereschi} for all DFT-MD simulations. For the conductivity calculations, at least ten uncorrelated snapshots are taken from the DFT-MD simulation and reevaluated using a more accurate energy convergence criterion and a 2x2x2 Monkhorst-Pack grid~\cite{Monkhorst}. Additionally, these snapshots were evaluated using the hybrid XC-functional of Heyd, Scuseria and Enzerhof~(HSE)~\cite{HSE1,HSE2}. For these calculations, however, only the Baldereschi mean-value point was considered due to the higher computational demand of HSE calculations. The electrical conductivity is computed from these simulations via the Kubo-Greenwood fomula~\cite{Kubo_stat,Greenwood}, using the eigenstates and eigenenergies of the Kohn-Sham orbitals (for details, see Ref.~\cite{Holst}). We employ a complex shift of $0.1$ in the Kramers-Kronig transformation.
We have carefully checked the convergence of our results with regard to plane wave energy cutoff, length of the time step, number of particles and Brillouin zone sampling.
Additionally, we compute electronic transport properties using time-dependent DFT (TD-DFT) in the linear response regime~\cite{Gross1985}, which is based on the linear density-density response of the electronic system to an external, time-dependent perturbation.

Furthermore, we train high-dimensional neural network interatomic potentials to reproduce the DFT forces and energies, enabling us to perform neural-network-driven molecular dynamics (NN-MD) simulations with upto 32~000 copper atoms. Separate neural network potentials are trained for each condition due to the large temperature range covered in this study. Each neural network is trained on at least 4000 configurations randomly sampled from DFT-MD simulations that span at least 40 000 time steps. These simulations include a long (at least 20 000 time steps) simulation at the desired conditions and shorter (at least 5000 time steps) simulations at slightly higher and lower density and temperature than the considered conditions. The higher particle number improves the resolution of the phonon dispersion in the solid due to the larger number of available wave vectors, and it enables access to the hydrodynamic limit in the liquid regime. We use the implementation in the n2p2 software package~\cite{n2p2,Morawietz2016,Singraber2019}, which employs Behler-Parrinello symmetry functions to describe the environment of each copper atom and subsequently passes these symmetry functions to the input layer of the neural network. A Kalman filter is used to adjust the weights and biases of the neural network during training. We use two hidden layers with 40 nodes each, and set the cutoff radius between 6~\AA{} at ambient conditions and 4~\AA{} at the highest pressure condition along the Hugoniot. The environment of each atom is described by 13 radial symmetry functions and 12 narrow angular symmetry functions chosen according to the scheme described in Ref.~\cite{Gastegger2018}. The remaining parameters are set to their default values. The trained potential is subsequently used in conjunction with the LAMMPS molecular dynamics simulation code~\cite{Thompson2022} to produce the MD simulations. The temperature in the NN-MD simulations is also controlled using a Nos\'e{}-Hoover thermostat.

\section{Results for solid copper at ambient conditions}
\label{sec:solid-metal}

\begin{figure}[htb]
\center{\includegraphics[angle=0,width=1.0\linewidth]{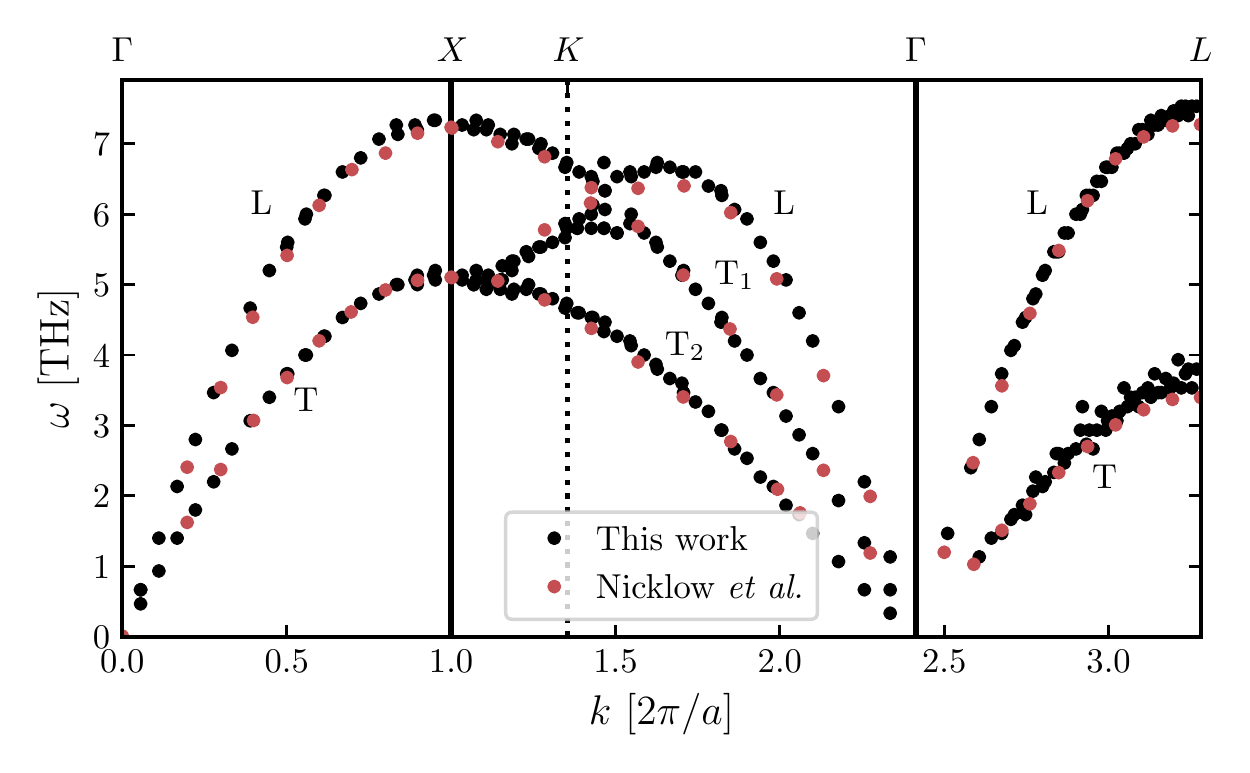}}
\caption{The phonon dispersion of solid copper at $\rho = 8.94$~g/cm$^3$ and $T=303$~K extracted from the longitudinal and transverse current-current correlation spectrum. The x-axis is scaled by the lattice constant $a$ of ambient copper. Experimental data from Nicklow~\textit{et al.} is given as a reference~\cite{Nicklow1967}.}
\label{fig:phonon}
\end{figure}

First, we test our simulations against known results for ambient solid copper at the density $\rho=8.94$~g/cm$^3$ and temperature $T=303$~K. While the DSF described in Eq.~\eqref{eq_Sii_WK} can be averaged over all wave vectors with equal magnitude in liquid and warm dense copper, the orientation of $\vec{k}$ relative to the crystallographic axes is relevant in the solid phase.
In Fig.~\ref{fig:phonon}, we show the phonon dispersion of solid copper at ambient conditions along a high-symmetry path through the first Brillouin zone. The phonon positions are determined from the transverse and longitudinal current-current correlation spectrum $J_\mathrm{t} (k, \omega)$ and $J_\mathrm{l} (k, \omega)$ (see Ref.~\cite{Schoerner2022} for details) of the NN-MD simulations by the peak finding routine \textit{find\_peaks} implemented in the SciPy library for scientific computing in python~\cite{scipy2020}. This analysis of phonon modes is fully dynamic and does not require a harmonic or quasi-harmonic oscillator model.

\begin{figure}[htb]
\center{\includegraphics[angle=0,width=1.0\linewidth]{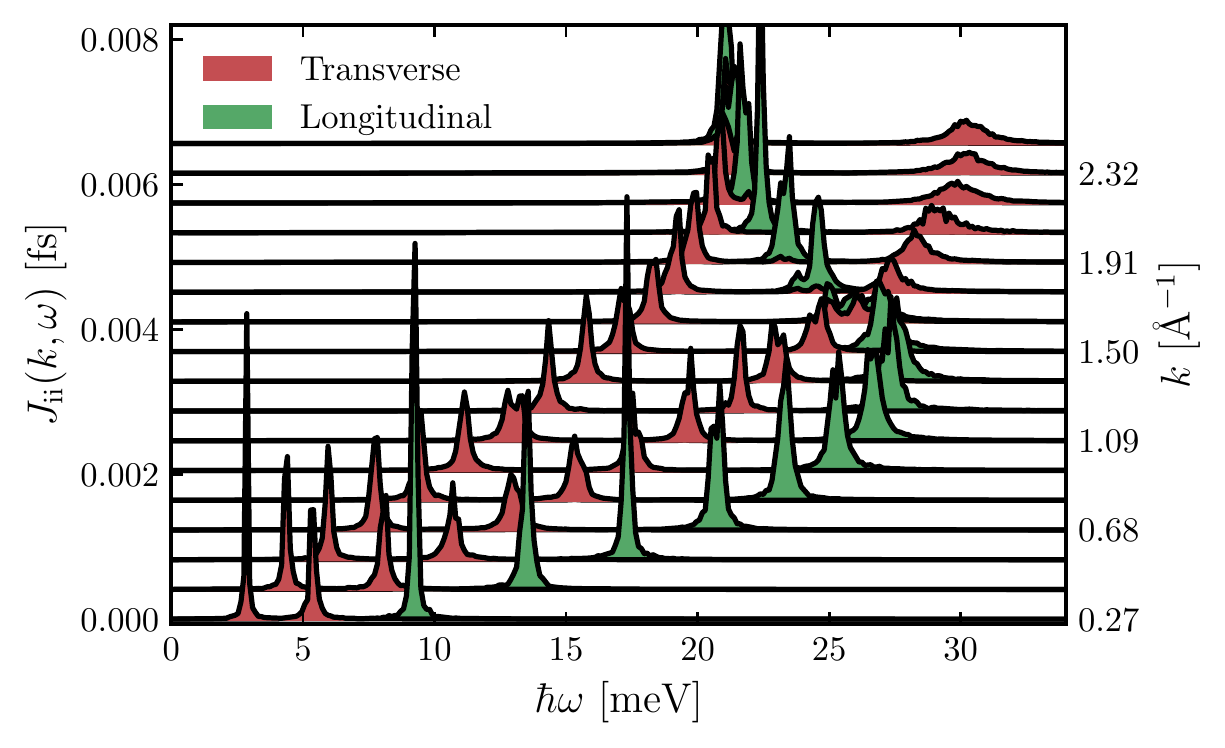}}
\caption{The current-current correlation spectrum of solid copper at $\rho = 8.94$~g/cm$^3$ and $T=303$~K along the high-symmetry path $\Gamma-K-X$. The transverse part is colored red, while the longitudinal part is colored green.}
\label{fig:phonon3D}
\end{figure}

The observed agreement with experimental data by Nicklow~\textit{et~al.}~\cite{Nicklow1967} is very good, indicating that the lattice dynamics in solid copper are well described by our simulations. We show an example of the transverse and longitudinal current-current correlation spectrum along the high-symmetry path $\Gamma-K-X$ in Fig.~\ref{fig:phonon3D}. The contributions due to longitudinal density oscillations are colored green and the the correlation spectrum of transverse currents is colored red.

\begin{figure}[htb]
\center{\includegraphics[angle=0,width=1.0\linewidth]{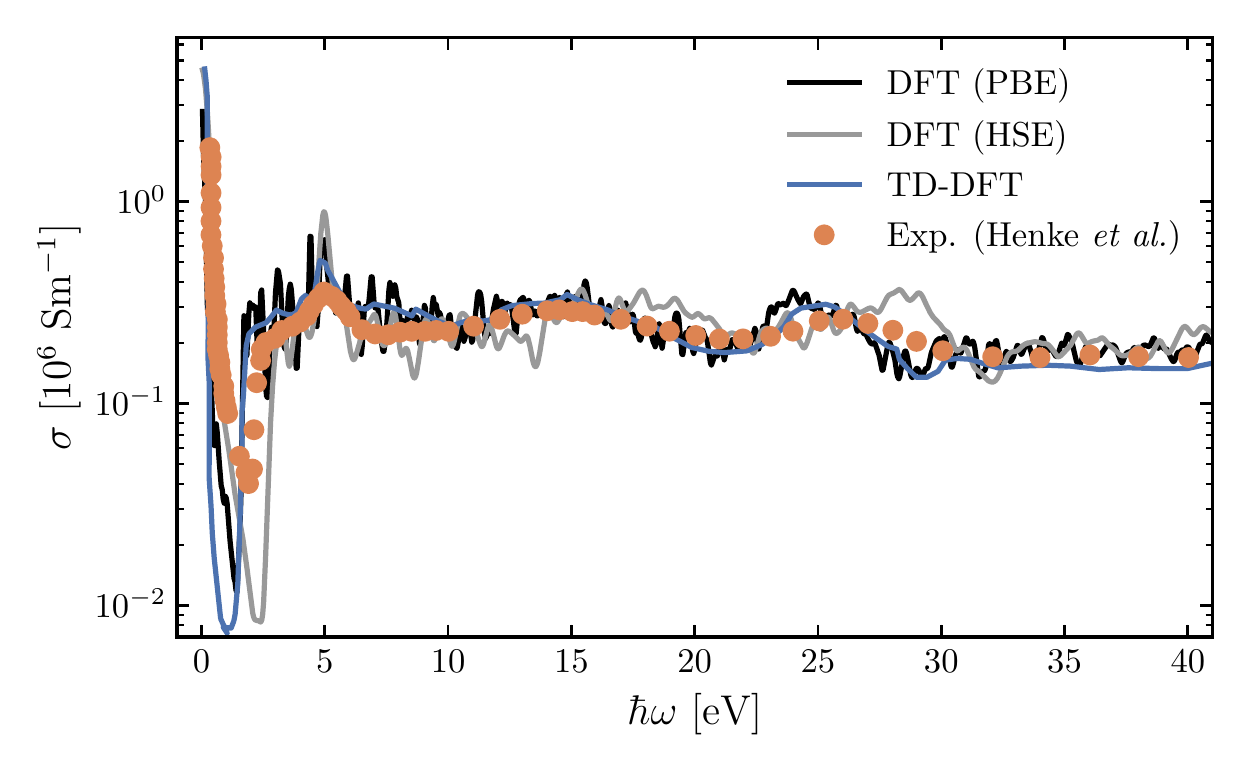}}
\caption{Dynamic electrical conductivity of solid copper at ambient conditions computed from a DFT simulation with 125 atoms using the Kubo-Greenwood formula and results of TD-DFT linear response calculations with 4 atoms and the adiabatic local density approximation. The black line indicates results achieved with the PBE XC-functional, while the grey line represents results using the HSE XC-functional. Measurements of Henke~\textit{et al.}~\cite{Henke} are given as reference.}
\label{fig:solid-cond}
\end{figure}

Furthermore, we investigate the dynamic electrical conductivity using the Kubo-Greenwood formula~\cite{Kubo_stat,Greenwood}. Results for PBE and HSE calculations are shown in Fig.~\ref{fig:solid-cond} compared to an experimental result by Henke \textit{et al.}~\cite{Henke} and predictions from TD-DFT linear response calculations. Experimentally, Henke~\textit{et al.} measured the absorption coefficient $\alpha(\omega)$ of ambient copper which can be translated to the real part of the dynamic electrical conductivity by Kramers-Kronig relations.
We also compute the dynamic electrical conductivity using linear response TD-DFT. Here, we compute the density-density response function due to an external perturbation potential within a simulation cell containing 4 atoms. The effect of electron-electron interactions is incorporated using the random-phase approximation, which accounts for local-field effects from the Coulomb interaction but neglects exchange-correlation. From the response function, transport properties, such as the electrical conductivity or the DSF, are extracted by aid of the fluctuation-dissipation theorem. TD-DFT has been used to compute XRTS spectra using both the real-time~\cite{Baczewski} and linear response formalisms~\cite{Ramakrishna2021}.
The agreement in Fig.~\ref{fig:solid-cond} is good, while the theoretical models predict a significantly lower conductivity at $\approx 2$~eV. Also, the measurements are lower than the theory predictions at $\approx 7$~eV and the feature at $\approx 25$~eV, which is observed by theory and experiment, occurs at lower frequencies in the simulations. This shift corresponds to shifted energy states relative to the continuum of states. The DFT-MD simulation predicts the energy states responsible for the observed feature at higher energies than the experiment indicates. This underestimation of energy gaps is a well known problem of DFT with commonly used XC-functionals like PBE~\cite{PBE}. Better agreement with experimental observations can be achieved using the hybrid functional HSE~\cite{HSE1, HSE2} as shown, e.g., for aluminum~\cite{Witte2018}. However, in this case, the HSE calculations overestimate the position of the feature, shifting it to $\approx 28$~eV.

\section{Results for liquid copper at ambient pressure}
\label{sec:liquid-metal}

\begin{figure}[htb]
\center{\includegraphics[angle=0,width=1.0\linewidth]{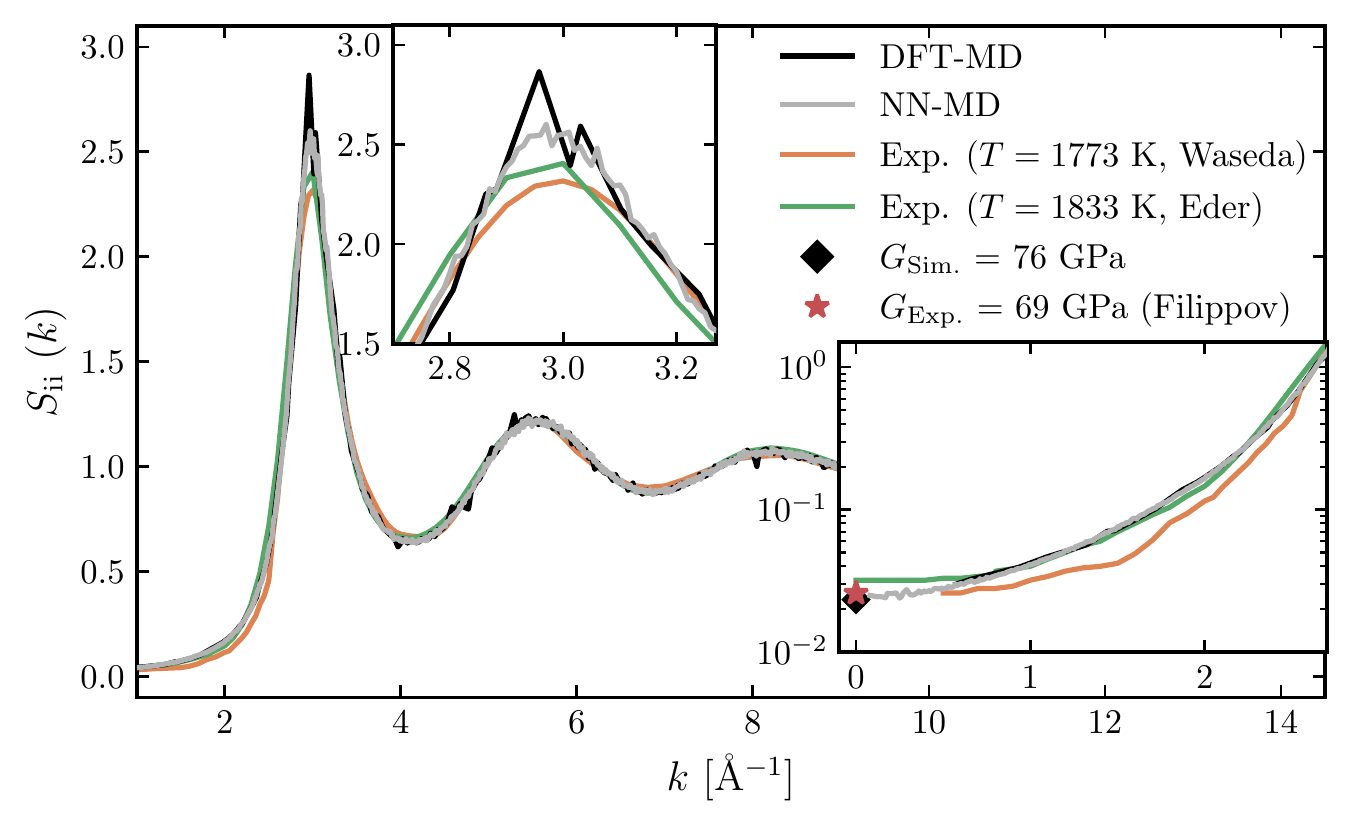}}
\caption{Static structure factor of liquid copper at ${\rho = 7.69}$~g/cm$^3$ and $T=1773$~K computed from a DFT-MD simulation with 125 atoms and a NN-MD simulation with 32~000~atoms. Two experimental results from x-ray diffraction (Waseda and Ohtani~\cite{Waseda}) and neutron diffraction (Eder \textit{et al.}~\cite{Eder1980}) are shown as reference. The right inset zooms in on the long-wavelength behavior and shows the $k \rightarrow 0$ prediction by DFT-MD and an experimental result of liquid copper near the melting point by Filippov~\cite{Filippov1966}. The top inset zooms in on the behavior around the first correlation peak.}
\label{fig:statS1770}
\end{figure}

As another test of the method we perform simulations of liquid copper at $\rho = 7.69$~g/cm$^3$ and $T=1773$~K in order to compare to experimental data by Waseda and Ohtani~\cite{Waseda} who performed x-ray diffraction experiments at these conditions. We additionally compare to neutron diffraction data by Eder~\textit{et al.}~\cite{Eder1980}.
The atomic form factor must be regarded if the ion dynamics are to be extracted from x-ray scattering. Waseda and Ohtani used form factors computed from relativistic Dirac-Slater wave functions~\cite{Cromer_Waber} with anomalous dispersion corrections~\cite{Cromer}, while for neutron diffraction merely the multiple scattering in the sample must be accounted for~\cite{Eder1980}. Fig.~\ref{fig:statS1770} shows static ion structure factors calculated from a DFT-MD simulation with 125 atoms and a NN-MD simulation with 32~000~atoms.

The static structure factors inferred from Waseda and Ohtani ($T=1773$~K) and from the diffraction experiments of Eder \textit{et al.} ($T=1883$~K) are displayed for comparison. The general agreement is good while there are noticeable differences in the first correlation peak at around $3\, \mathrm{\AA}^{-1}$ and for the low-$k$ limit, displayed in the insets of Fig.~\ref{fig:statS1770}. The first correlation peak is characterized by the length scale at which the minimum of the interatomic potential occurs. Here, longer simulations lead to better statistics, which better resolves the dynamics in this area, leading to a lowering of the peaks. Additionally, smaller simulation boxes artificially enhance the near-field order that is induced by the minimum of the interatomic potential.
Experimentally, it is influenced by the angular resolution of the detector and spectral width of the light/neutron source. Remarkably, the first correlation peak of Eder~\textit{et al.} lies higher than that of Waseda and Ohtani, although higher temperatures generally lead to diminishing correlations, corresponding to a lower correlation peak.

The low-$k$ limit is of interest because the value for $k \to 0$ can be determined by the isothermal compressibility via Eq.~\eqref{eq-Sk0} which is accessible through DFT-MD simulations via Eq.~\eqref{eq_th_compress}. We perform additional simulations at 5\% higher and lower densities in order to evaluate the derivative.
The limit determined this way is indicated by the black diamond in the inset, while the violet asterisk is determined from inserting an experimental compressibility (inferred from the speed of sound)~\cite{Filippov1966} into Eq.~\eqref{eq-Sk0}. Eder \textit{et al.} artificially extended their structure factor for $k$-values smaller than $0.5\,\mathrm{\AA}^{-1}$ to match the value computed from the compressibility near the melting point by Filippov~\textit{et al.}~\cite{Filippov1966}, also used for the violet asterisk, and a density determined by Cahill and Kirshenbaum~\cite{Cahill}.
It approaches a higher value than the two indicated limits due to the higher temperature of the experiment (see Eq.~\eqref{eq-Sk0}). The DFT-MD simulation with 125 atoms allows access to $k$-values down to $0.57$~\AA$^{-1}$, which indicates that the trend of $S_\mathrm{ii}(k)$ agrees qualitatively with the known limit at $k \to 0$, while no definite conclusion on the agreement can be made without further investigation with more atoms. With the larger NN-MD simulations, it is apparent that $S_\mathrm{ii}(k)$ also agrees quantitively with the limit computed through~\eqref{eq-Sk0}. For wave vectors $k$ between $0.8\,\mathrm{\AA}^{-1}$ and $2.5\,\mathrm{\AA}^{-1}$, observations differ between Eder~\textit{et al.} and Waseda and Ohtani. The DFT-MD simulations agree well with the former while the latter is significantly lower in that region (see inset in the lower right corner of Fig.~\ref{fig:statS1770}). A reason for this difference could be the form factor which must be additionally considered for x-ray diffraction experiments.

\begin{figure}[!thb]
\center{\includegraphics[angle=0,width=1.0\linewidth]{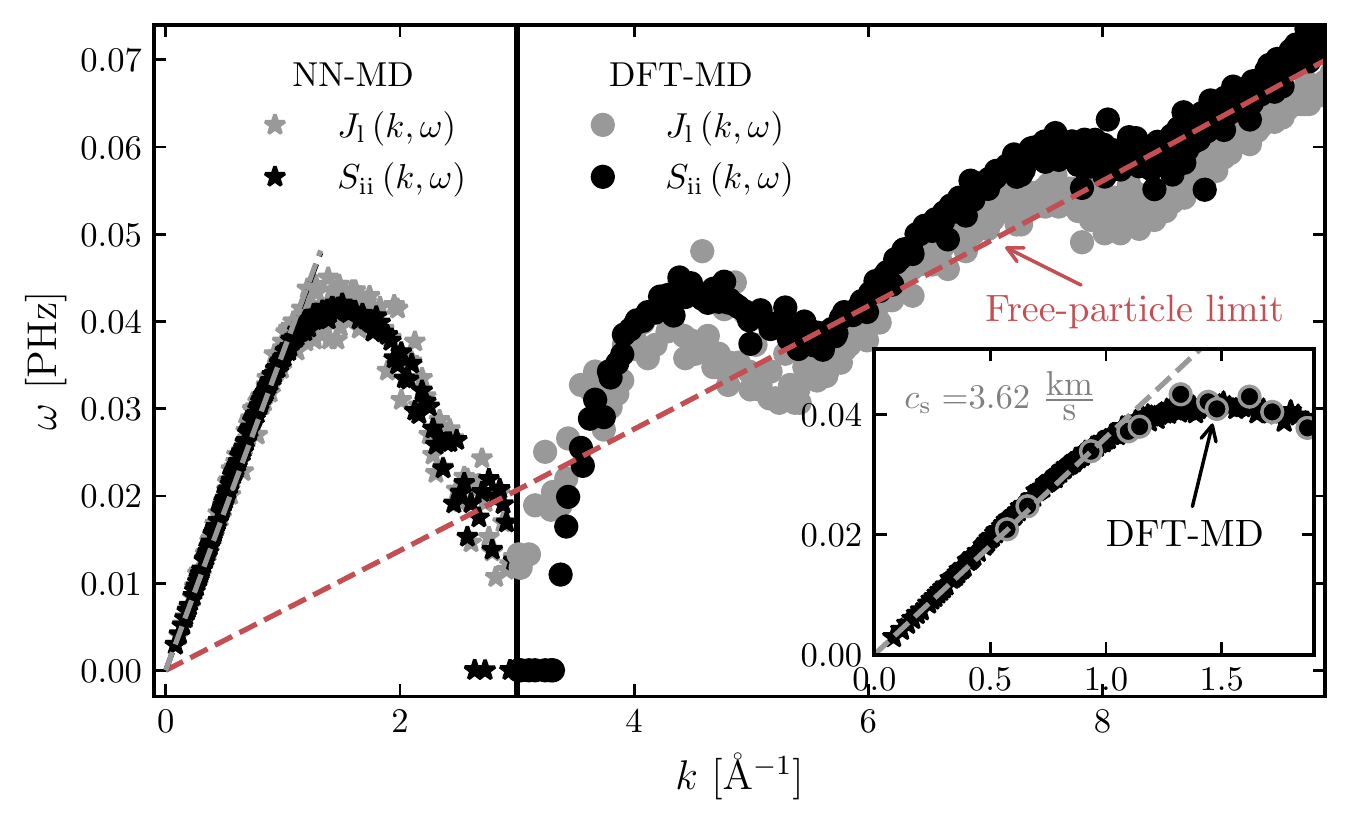}}
	\caption{Peak position of the ion acoustic mode taken from the DSF and from the longitudinal current-current correlation $J_\mathrm{l} \left(k,\,\omega\right)$ (see Eq.~\eqref{eq_current_correl}) dependent on the $k$-value. For wave numbers below the first correlation peak results from the NN-MD simulation are shown, while all other results are taken from the DFT-MD simulation. A best linear fit for the low $k$ limit of the ion acoustic mode of the DSF is indicated and the corresponding adiabatic speed of sound is presented. The inset shows a comparison between DFT-MD and NN-MD results for small $k$.}
\label{fig:peakpos1770}
\end{figure}

Furthermore, we compute the DSF~$S_\mathrm{ii} (k, \, \omega)$ and the closely related longitudinal current-current correlation spectrum~$J_\mathrm{l} (k, \, \omega)$. We extract various properties of the different contributing modes by fitting to a generalized collective modes approach~\cite{Bryk2012,Wax2013} with one diffusive and one propagating mode, see Ref.~\cite{Schoerner2022} for details.
Fig.~\ref{fig:peakpos1770} shows the peak positions of the ion acoustic mode extracted from the DSF as a function of the wave vector $k$. This can be considered as the dispersion relation of the ion acoustic mode. Also shown in this figure are the peak positions of the longitudinal current correlation given by Eq.~\eqref{eq_current_correl} which describes collective excitations via currents~\cite{Zwanzig} and determines the apparent speed of sound $c_\mathrm{l}$~\cite{Balucani}. The DSF, on the other hand, can be used to extract the adiabatic speed of sound $c_\mathrm{s}$ in the hydrodynamic limit~\cite{Hansen2006}. Both of these quantities are indicated in Fig.~\ref{fig:peakpos1770}, as well as the free-particle limit of a non interacting classical system. In this limit, the peak position of $J_\mathrm{l}$ is determined by the non collective mode which is described by
\begin{equation}
\label{eq:freeparticle}
\omega_J (\vec{k}) = \sqrt{ \frac{2}{m_\mathrm{i} \beta}} \; |\vec{k}| \, ,
\end{equation}
see Ref.~\cite{Scopigno2}. The different physical regimes that correspond to low- and high-$k$-values were discussed in detail recently in Ref.~\cite{Schoerner2022} for warm dense aluminum. As the ions behave like free particles in the high-$k$ limit, their dispersion is described by the free-particle limit, as can be seen from Fig.~\ref{fig:peakpos1770}.

\begin{figure}[htb]
\center{\includegraphics[angle=0,width=1.0\linewidth]{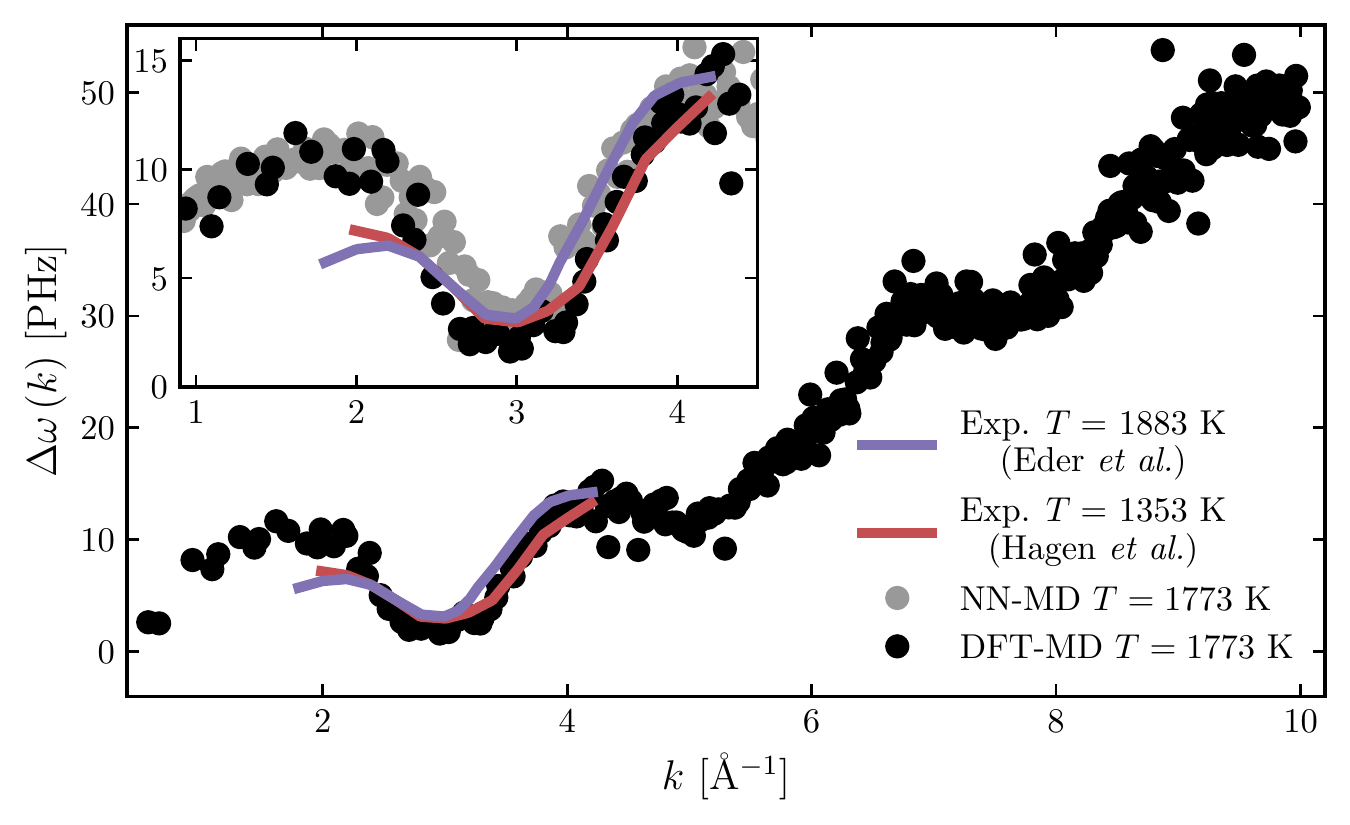}}
\caption{Peak width of the thermal mode of the ion-ion DSF dependent on the $k$-value. Experimental results from Hagen~\textit{et al.}~\cite{Hagen} and data calculated from experimentally determined static ion structure factors by Eder~\textit{et al}.~\cite{Eder1980} are given as comparison. The inset shows a comparison between the results of the DFT-MD~(125~atoms) and NN-MD~(32~000~atoms) simulations in the region where experimental data is available.}
\label{fig:peakwidth1770}
\end{figure}

Another feature that can be extracted from the DSF is the width of the diffusive thermal mode represented in Fig.~\ref{fig:peakwidth1770}.
It corresponds to the random thermal movement of the ions and its shape is connected to how much energy can be coupled to this mode. The wider the peak, the higher the energy transfer to an individual ion can be. The magnitude of the peak is determined by the static structure factor which accounts for how many ions are present on the length scale defined by $k$. Therefore, the width and the height of the peak determine the energy that can be transferred to the thermal mode. In Fig.~\ref{fig:peakwidth1770}, the widths for DFT-MD simulations with 125~atoms and NN-MD simulations with 32~000~atoms are shown and compared to an experimental result from inelastic neutron scattering by Hagen~\textit{et al.}~\cite{Hagen} and a result inferred from neutron scattering by Eder~\textit{et al.}~\cite{Eder1980}. While the former represents a direct measurement, the results by Eder~\textit{et al.} are calculated from the static structure factor in Fig.~\ref{fig:statS1770}. They use a simple model~\cite{Copley} with one free parameter to artificially introduce the dynamics. The ab-initio DFT-MD approach describes the dynamics in a more consistent way than the model chosen by Eder~\textit{et al.} Since their static structure factor agreed well with DFT-MD simulations (see Fig.~\ref{fig:statS1770}), the observed deviations are attributed to the artificially introduced dynamics.

\section{The principal Hugoniot curve}
\label{sec:hugo}

The Hugoniot equation~\eqref{eq_hugoniot} is dependent on pressure $P$, density $\rho$, and specific internal energy $\epsilon$. For a given temperature, the equation of state (EOS) defines the values for $P$, $\rho$, and $\epsilon$ that satisfy the Hugoniot equation. We compute 16 isotherms ranging from 1700~K upto 60~000~K, with four to five different densities per isotherm.

\begin{figure}[htb]
\center{\includegraphics[angle=0,width=1.0\linewidth]{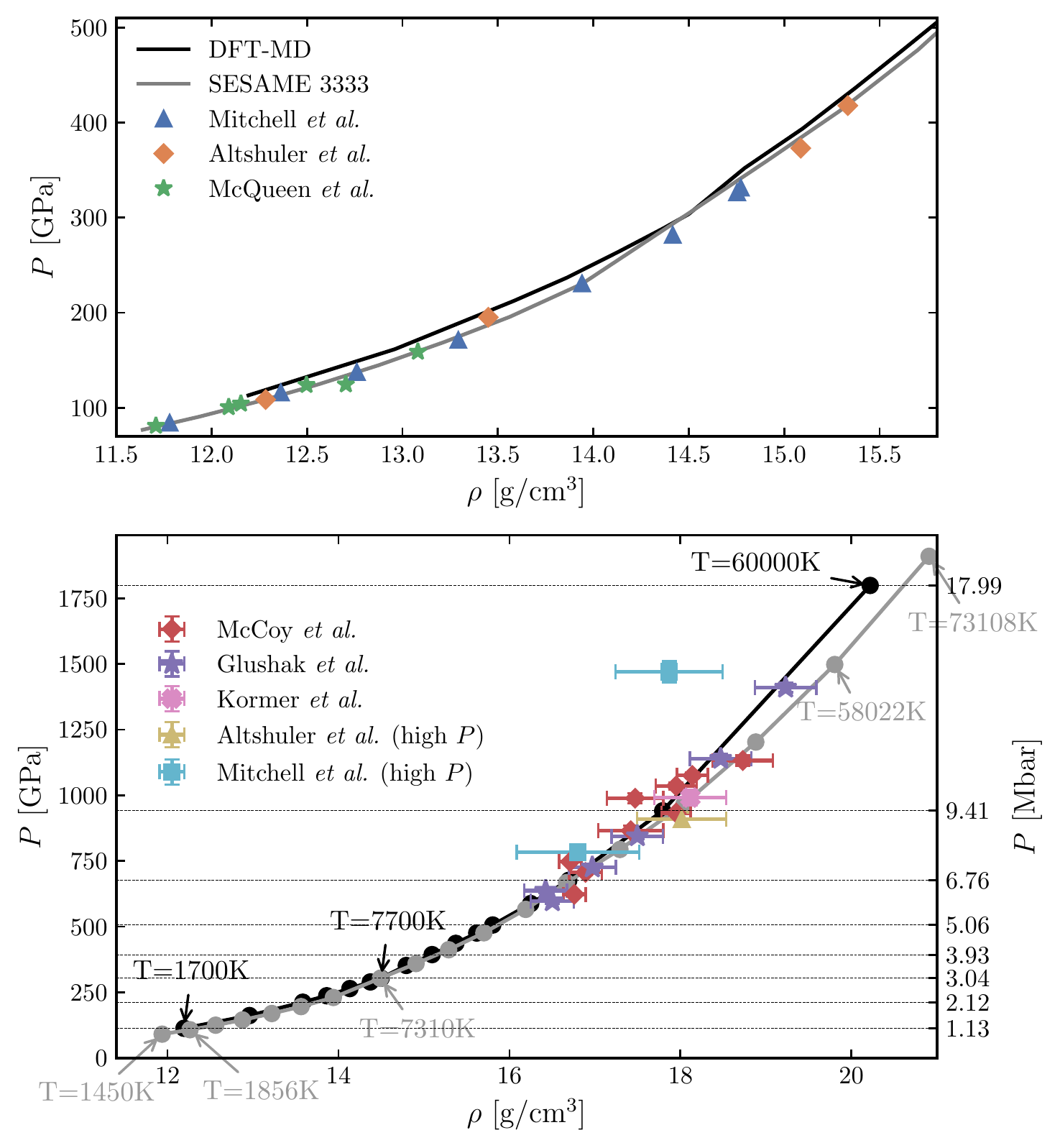}}
\caption{Lower panel: Hugoniot curve for copper in the $P$-$\rho$ plane inferred from isotherms calculated using DFT-MD (black) and isotherms from the SESAME 3333 table~\cite{Trainor1983}~(grey). The temperatures for some data points are annotated. High pressure experimental values by McCoy~\textit{et~al.}~\cite{McCoy2017}, Glushak~\textit{et~al.}~\cite{Glushak1989}, Kormer~\textit{et~al.}~\cite{Kormer1962}, Altshuler~\textit{et~al.}~\cite{Altshuler1962} and Mitchell~\textit{et~al.}~\cite{Mitchell1991} are shown. The dotted horizontal lines indicate the conditions we choose for furhter investigation. More information on the conditions is given in Tab.~\ref{tab:hugo_cond}. \\
Upper panel: Zoomed in view of the low pressure range, where experimental data by Mitchell~\textit{et~al.}~\cite{Mitchell}, Altshuler~\textit{et~al.}~\cite{Altshuler}, McQueen~\textit{et~al.}~\cite{McQueen} are shown.}
\label{fig:hugo_exp}
\end{figure}

The pressure and internal energy are interpolated using cubic splines, and Eq.~\eqref{eq_hugoniot} is solved numerically to give the principal Hugoniot curve depicted in Fig.~\ref{fig:hugo_exp}. As a comparison, we compute the principal Hugoniot curve from the standard SESAME 3333 EOS table~\cite{Trainor1983}.
Fig.~\ref{fig:hugo_exp} illustrates the results obtained from DFT-MD isotherms and from SESAME isotherms in the pressure-density plane.
Experimentally, the Hugoniot curve of copper in the pressure-density plane has been constrained well for pressures up to $4$~Mbar (see Fig.~\ref{fig:hugo_exp}). For higher pressures, the spread of experimental results is significantly larger~\cite{McCoy2017} and experimental uncertainties increase.

\begin{table}[tb]
\caption{Conditions for compressed copper predicted from DFT-MD isotherms.}
\label{tab:hugo_cond}
\begin{ruledtabular}
\begin{tabular}{cccc}
$P$ [GPa]& $T$ [K] & $\rho$ [g/cm$^3$] & $u$ [kJ/g] \\
\hline
113  & $1700$  & $12.193$ & $-3.9774$ \\
162  & 3000    & 12.959 & $-2.8628$ \\
212  & $4600$  & $13.583$ & $-1.6052$ \\
237  & 5400    & 13.862   & $-0.9597$ \\
264  & 6300    & 14.134   & $-0.2339$ \\
289  & 7200    & 14.373   & 0.4534    \\
304  & $7700$  & $14.499$ & $0.8458$  \\
352  & 8200    & 14.793   & 2.1275    \\
393  & $9500$  & $15.095$ & $3.3089$  \\
436  & 10900   & 15.373   & 4.5471    \\
476  & 12400   & 15.617   & 5.7083    \\
506  & $13500$ & $15.805$ & $6.6302$  \\
586  & 16600   & 16.249   & 9.1182    \\
676  & $20000$ & $16.694$ & $11.8969$ \\
941  & $30000$ & $17.792$ & $20.5124$ \\
1799 & $60000$ & $20.220$ & $50.4514$
\end{tabular}
\end{ruledtabular}
\end{table}

\begin{figure}[htb]
\center{\includegraphics[angle=0,width=1.0\linewidth]{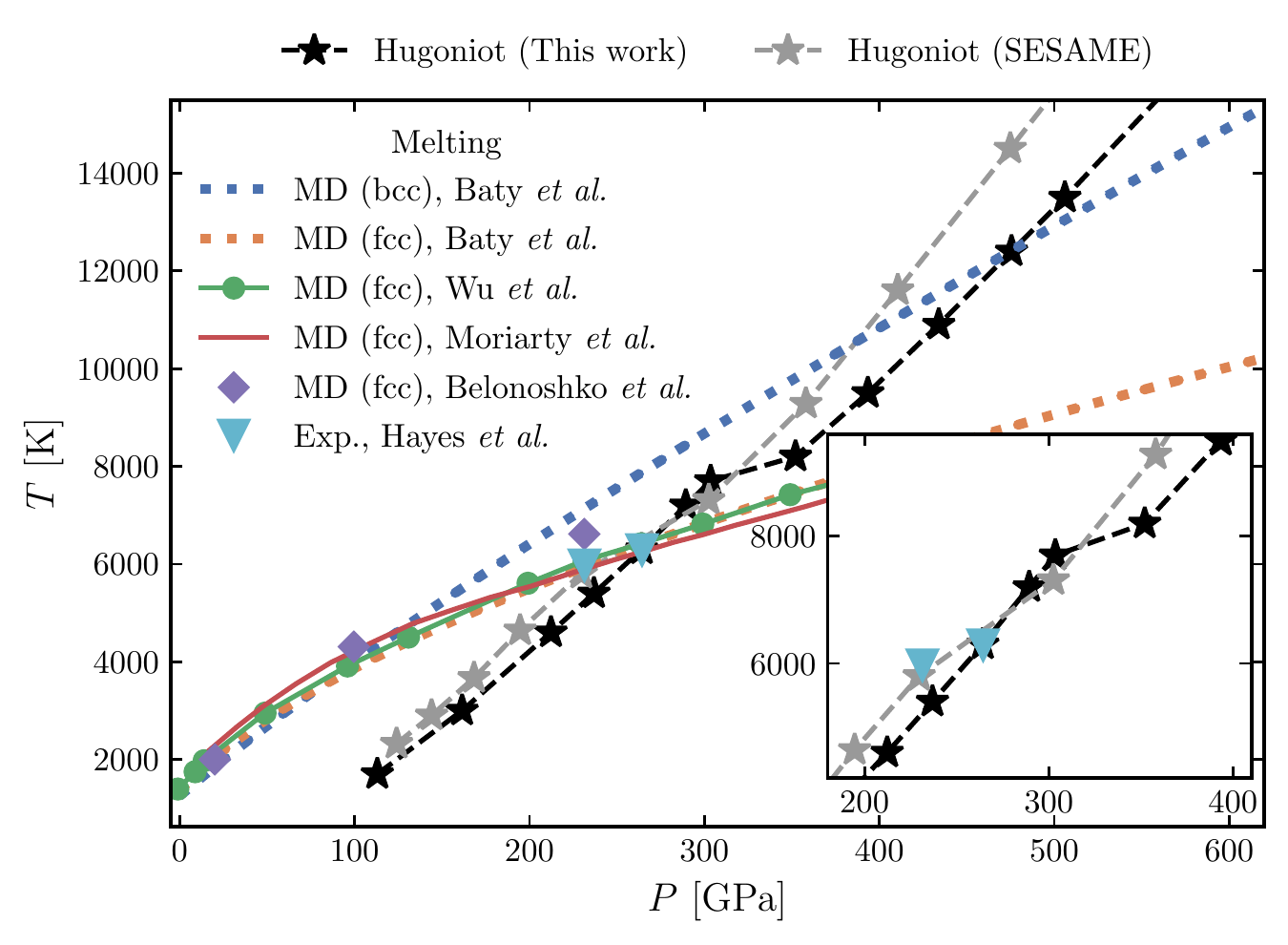}}
\caption{Hugoniot curve for copper in the $T$-$P$-plane inferred from isotherms calculated using DFT-MD (black) and isotherms from the SESAME 3333 table (grey). Melting lines determined by Wu \textit{et al.}~\cite{Wu2011}, Moriarty \textit{et al.}~\cite{Moriarty1987} and Belonoshko \textit{et al.}~\cite{Belonoshko2000} using different variations of MD simulations as well as experimental results by Hayes \textit{et al.}~\cite{Hayes2000} are indicated.}
\label{fig:hugo_melt}
\end{figure}

The SESAME EOS predicts consistently higher temperatures during the compression process. While the temperature difference at $\approx 1$~Mbar at roughly similar conditions is $\approx 150$~K, the difference becomes $\approx 3200$~K around $6.7$~Mbar. While the deviation in the pressure-density-plane is small and difficult to assess experimentally, the temperature difference is significant and could be used to test the respective EOS.
The agreement in Fig.~\ref{fig:hugo_exp} is best between $2.5$~Mbar and $3.5$~Mbar which is the region in which both EOS predict the melting point as can be seen in Fig.~\ref{fig:hugo_melt}.

Experimental results by Mitchell \textit{et al.}~\cite{Mitchell}, Altshuler \textit{et al.}~\cite{Altshuler} and McQueen \textit{et al.}~\cite{McQueen} are indicated in the upper panel of Fig.~\ref{fig:hugo_exp} and show better agreement with the SESAME data while the DFT-MD Hugoniot lies slightly higher in the pressure-density-plane.
The lower panel of Fig.~\ref{fig:hugo_exp} shows the available high-pressure Hugoniot compression data by McCoy~\textit{et~al.}~\cite{McCoy2017}, Glushak~\textit{et~al.}~\cite{Glushak1989}, Kormer~\textit{et~al.}~\cite{Kormer1962}, Altshuler~\textit{et~al.}~\cite{Altshuler1962} and Mitchell~\textit{et~al.}~\cite{Mitchell1991}. In this regime, both EOS studied here are compatible with the experimental data due to their large experimental uncertainties. The only exception is the high-pressure point by Mitchell~\textit{et~al.} around 1500~GPa, which agrees with neither of the theoretical predictions.

In the temperature-pressure plane, the melting point along the Hugoniot is easily identifiable by a kink which is due to the latent heat needed for the phase transition. From the inset in Fig.~\ref{fig:hugo_melt}, the melting point as predicted by SESAME lies between $2.3$~Mbar and $3$~Mbar, and the melting point predicted by DFT-MD calculations lies between $3$~Mbar and $3.5$~Mbar. While the MD simulations by Wu~\textit{et al.}~\cite{Wu2011} and Moriarty~\textit{et al.}~\cite{Moriarty1987}, as well as experiments by Hayes~\textit{et al.}~\cite{Hayes2000} agree roughly with the SESAME results, the two-phase MD simulations by Belonoshko~\textit{et al.}~\cite{Belonoshko2000} display a trend that tends towards the melting point predicted by DFT-MD simulations. However, their calculations did not cover the pressure range in question. A recent study by Baty~\textit{et~al.}~\cite{Baty2021} considers the melting line for the fcc structure of copper, as well as the melting line of the experimentally observed bcc phase, which lies above the fcc melting point along the Hugoniot. For the subsequent examination of the material along the principal Hugoniot curve, the conditions indicated by horizontal dashed lines in the upper panel of Fig.~\ref{fig:hugo_exp} were used to perform extended simulations.

\section{Ion dynamics of shock-compressed copper}
\label{sec:shock-copper}

\begin{figure}[!t]
\center{\includegraphics[angle=0,width=1.0\linewidth]{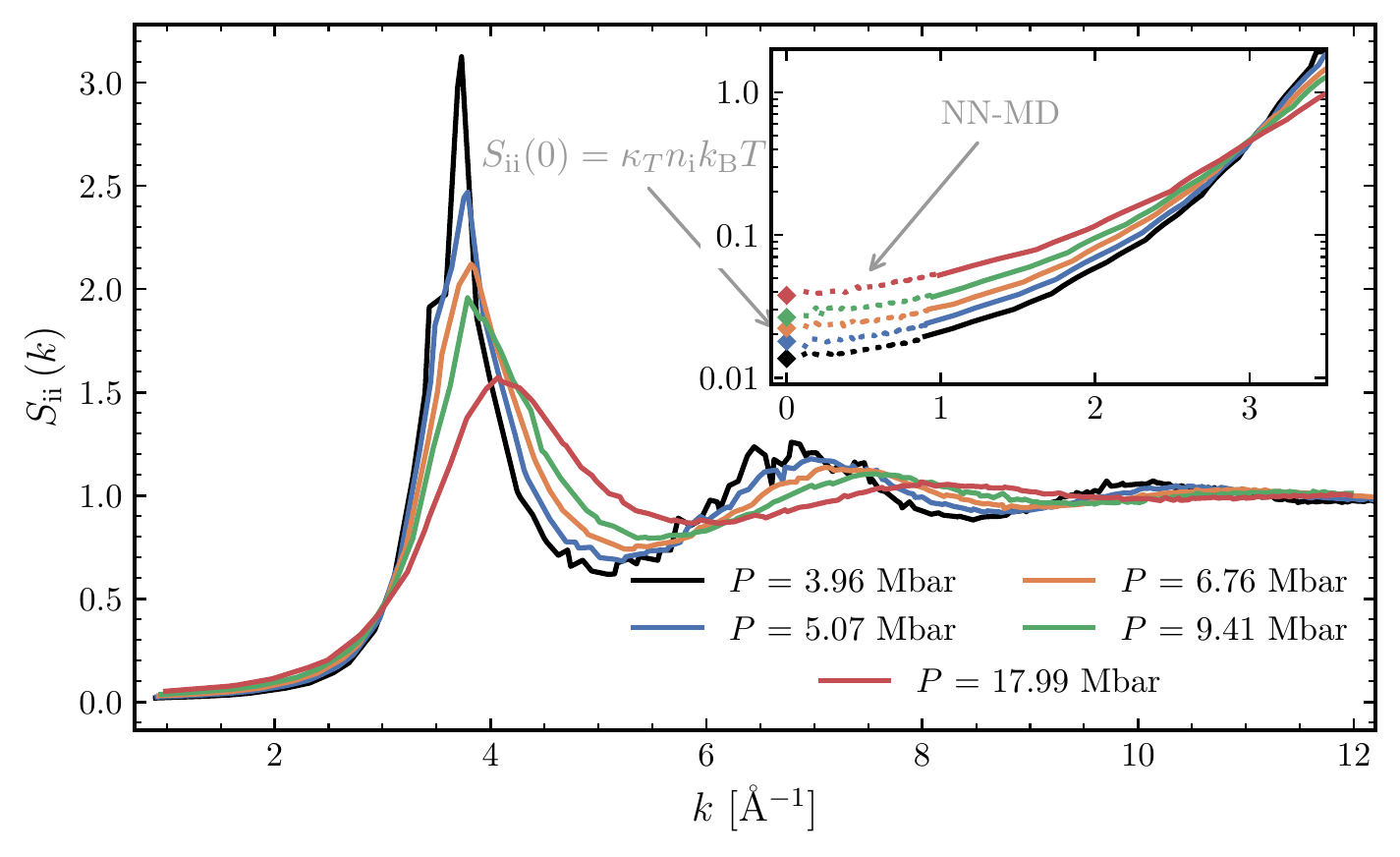}}
\caption{DFT-MD results of the static structure factor of liquid copper along the Hugoniot curve. In the inset the low $k$ behavior is shown on a log scale and the NN-MD results are shown for wave numbers that are not accessible to the DFT-MD results. The limits for $k \rightarrow 0$ are also indicated (see Eq.~\eqref{eq-Sk0}). Further information on the conditions is given in Fig.~\ref{fig:hugo_exp} and Tab.~\ref{tab:hugo_cond}.}
\label{fig-statS4568}
\end{figure}

The static ion structure factors in the liquid phase are displayed in Fig.~\ref{fig-statS4568}. The correlation peaks exhibit a shift to higher-$k$ values for increasing density. Shifts for constant pressure differences along the Hugoniot are expected to become smaller, as the density $\rho$ is a convex function of the pressure $P$ (as seen in Fig.~\ref{fig:hugo_exp}). A smoothing and lowering of the correlation peaks is also observable due to the lowering of the coupling parameter, caused by a combination of higher temperatures, higher densities and ionization. The limits for $k \to 0$ are also indicated by diamonds in the inset of Fig.~\ref{fig-statS4568}. The limits are computed via the compressibility using Eqs.~\eqref{eq_th_compress} and \eqref{eq-Sk0}, analogous to Sec.~\ref{sec:liquid-metal}. The lowest wave number available through the DFT-MD simulations is $0.9~\mathrm{\AA}^{-1}$, observed for the lowest density at $15.095$~g/cm$^{3}$, due to the small simulation boxes with 64 atoms. 

In order to test the computed limits, we perform NN-MD simulations with 32~000 atoms, enabling access to wave numbers down to $0.1~\mathrm{\AA}^{-1}$. The static structure factor at the additionally accessible wave numbers is given by the dashed lines in the inset of Fig.~\ref{fig-statS4568}, demonstrating good agreement with the DFT-MD data and the thermodynamically determined limit. The determined compressibilities will be used to compute the adiabatic speed of sound in the following.

\begin{figure}[!b]
\center{\includegraphics[angle=0,width=1.0\linewidth]{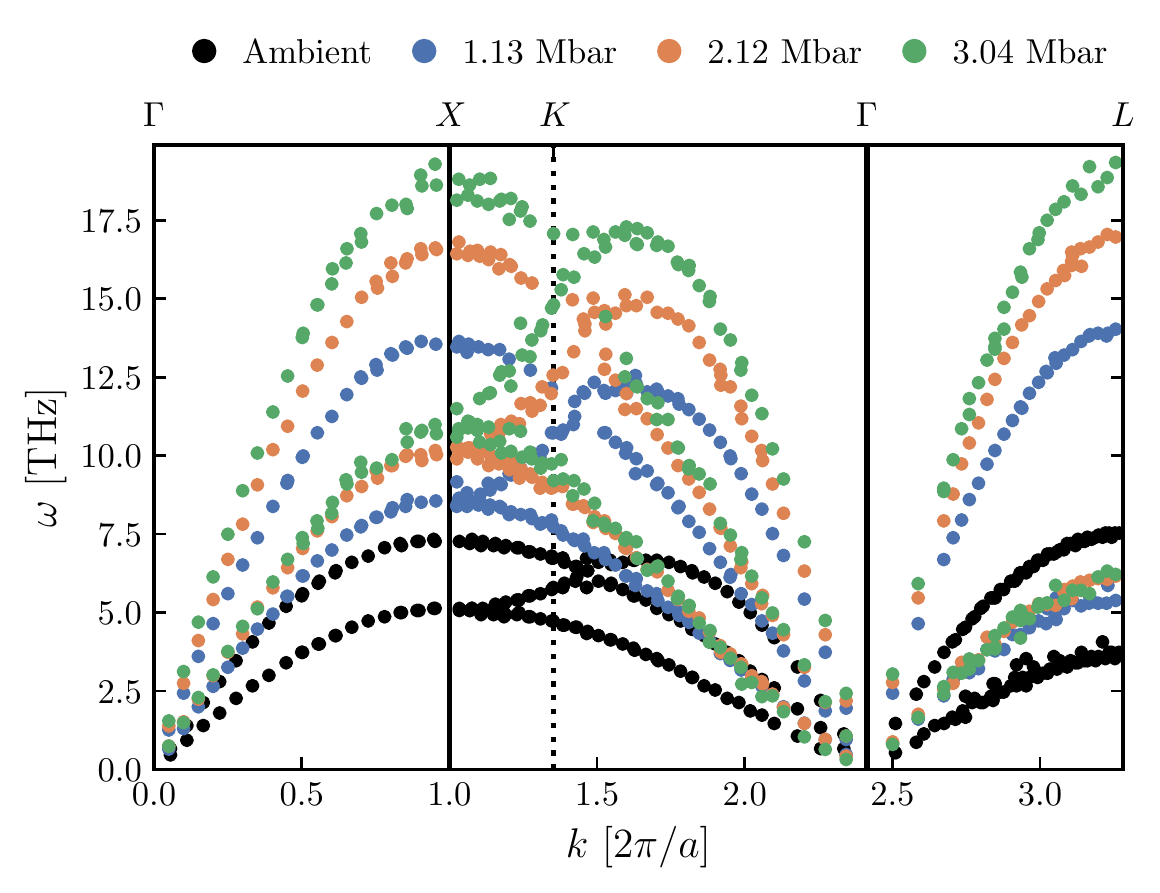}}
\caption{Phonon dispersion of solid copper along the principal Hugoniot curve along high-symmetry direction in the Brillouin zone. The x-axis is normalized by the lattice constant $a$ to make the results at different densities comparable. Further information on the conditions is given in Fig.~\ref{fig:hugo_exp} and Tab.~\ref{tab:hugo_cond}.}
\label{fig:phonon_hugo_solid}
\end{figure}

We study three points along the principal Hugoniot curve, where we expect solid conditions of copper according to Fig.~\ref{fig:hugo_exp} and Fig.~\ref{fig:hugo_melt}. The phonon dispersion of copper for ambient conditions and the Hugoniot conditions at $1.13$~Mbar, $2.12$~Mbar and $3.04$~Mbar are shown in Fig.~\ref{fig:phonon_hugo_solid}. While we only show the peak position here, a systematic broadening of the phonon modes is also observed, as expected due to the increasing temperature. Furthermore, a systematic hardening for most of the phonon branches and orientations can be observed. Along the $\Gamma-L$ direction, we observe a strong hardening of the longitudinal mode, while the transverse branch hardens significantly less than along the other shown orientations.

\begin{figure}[htb]
\center{\includegraphics[angle=0,width=1.0\linewidth]{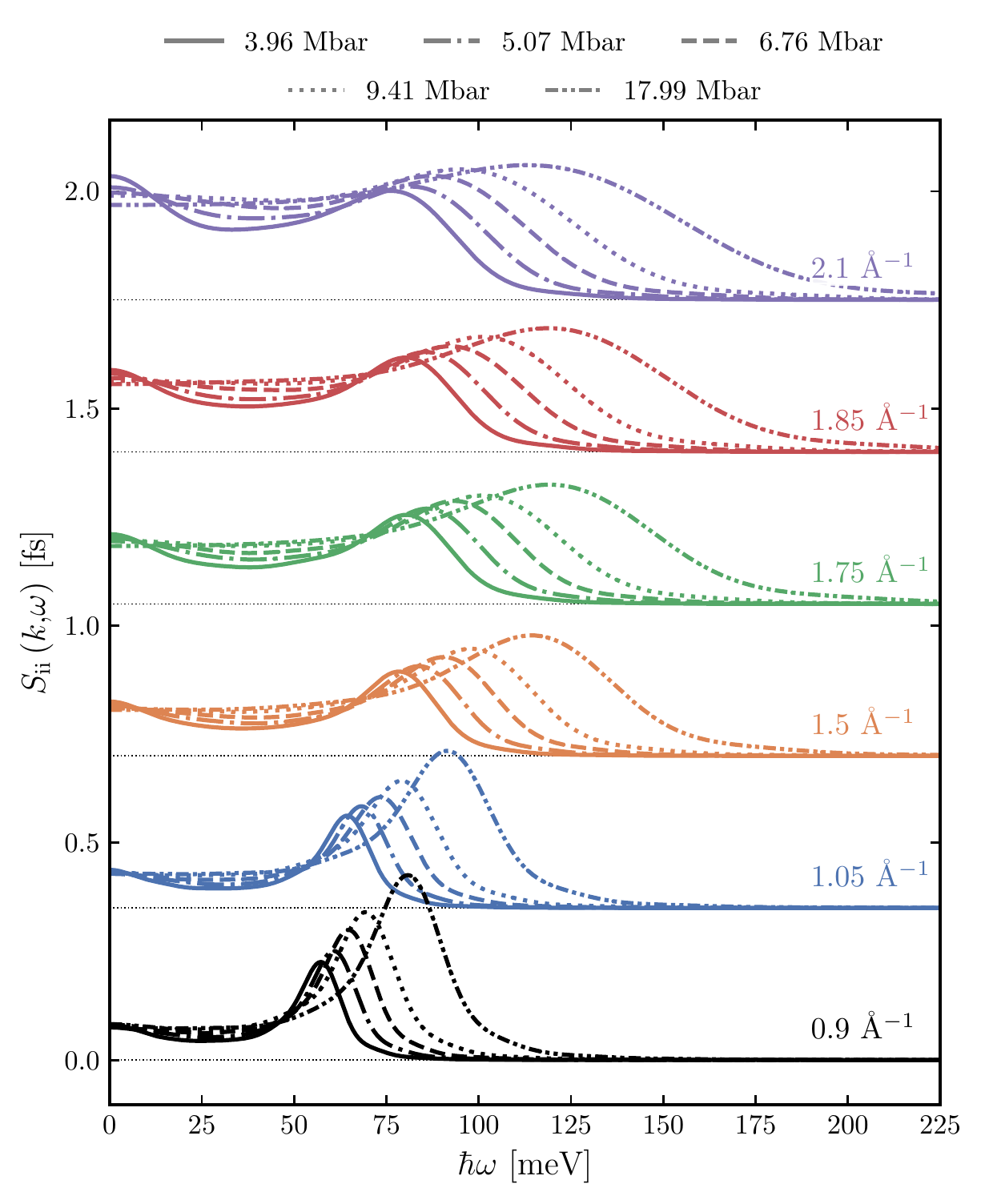}}
\caption{DSF of liquid copper along the principal Hugoniot curve computed from the NN-MD simulations. The curves are shifted by $0.35$~fs with respect to the next lower $k$-value for readability. Further information on the conditions is given in Fig.~\ref{fig:hugo_exp} and Tab.~\ref{tab:hugo_cond}.}
\label{fig:dynS_hugo_liquid}
\end{figure}

Once the copper melts, the effect of further compression on the ion acoustic mode can be investigated. Fig.~\ref{fig:dynS_hugo_liquid} shows the change of the dynamic ion structure factor at different $k$-values along the Hugoniot computed from the NN-MD simulations.

Since the simulations are performed at different densities, but with the same number of atoms, the size of the simulation box varies which leads to slightly different $k$-values in each case. However, the large-scale NN-MD simulations with $32~000$ atoms permit access to a dense $k$-grid in the considered range. Therefore, the $k$-values at all conditions are within $1$\% of the numbers given in Fig.~\ref{fig:dynS_hugo_liquid}. A similar study on aluminum~\cite{Schoerner2022} has shown that increased temperature leads to a more pronounced ion acoustic mode, but does not shift it to higher $\omega$. This effect is, therefore, attributed to the density increase along the principal Hugoniot curve. The more pronounced ion acoustic mode, as well as the generally elevated course of $S_\mathrm{ii} (k, \, \omega)$ for higher temperatures is in accord with the observation in Fig.~\ref{fig-statS4568} that the static structure factor $S_\mathrm{ii} (k)$ is greater at high temperatures than at low temperatures for $k$-values smaller than $2.8 \, \mathrm{\AA}^{-1}$.

\begin{figure}[htb]
\center{\includegraphics[angle=0,width=1.0\linewidth]{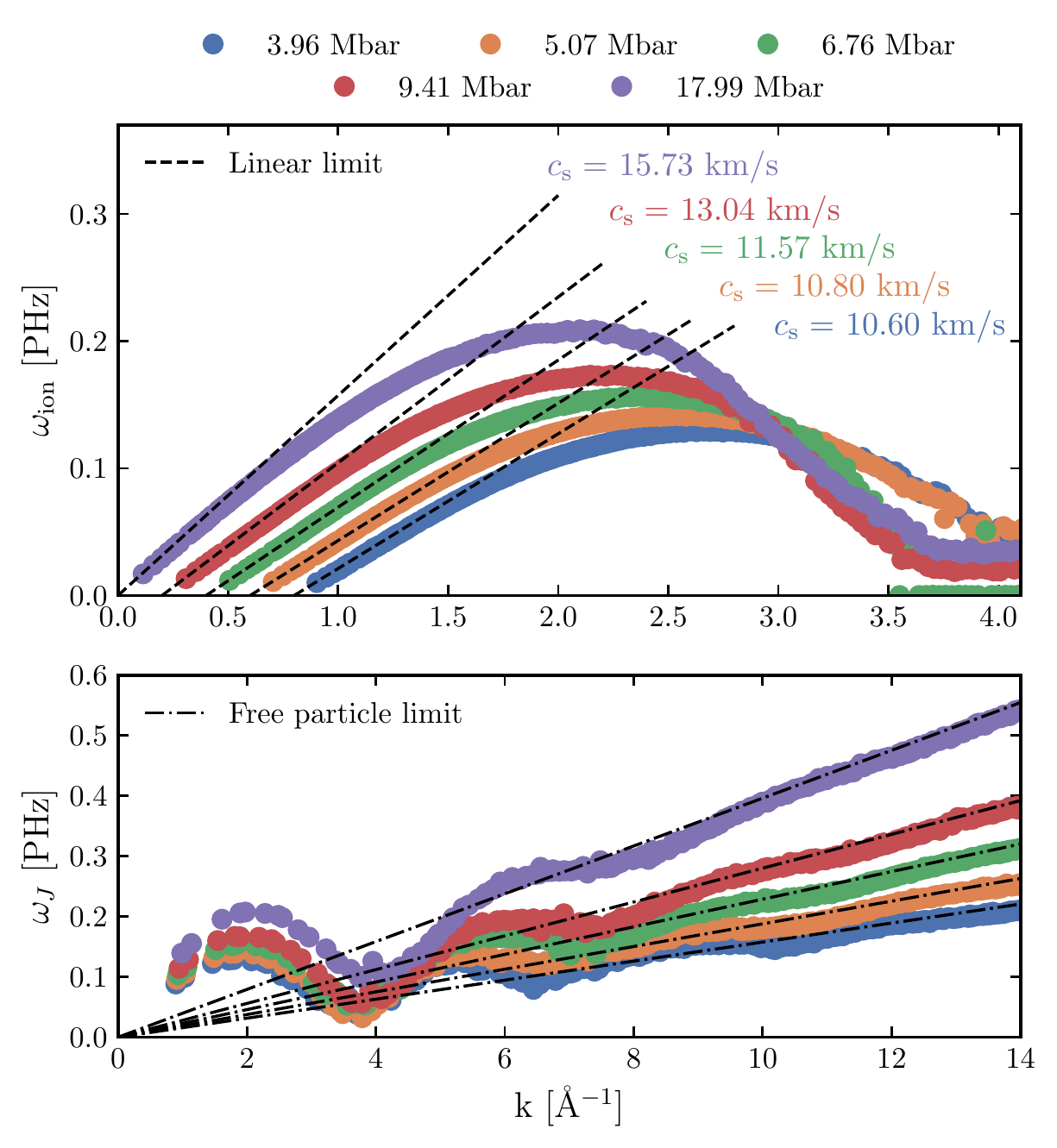}}
\caption{The upper panel shows the peak position of the ion acoustic mode $\omega_\mathrm{ion}$ taken from the DSF's in Fig.~\ref{fig:dynS_hugo_liquid} dependent on the $k$-value and a linear dispersion computed from equation~\eqref{eq:cs}. The corresponding adiabatic speeds of sound are presented.  The lower panel shows the dispersion of the longitudinal current-current correlation spectrum $J_\mathrm{l}$ extracted from the NN-MD simulations and the free-particle behavior. For readability, the curves in the upper panel are shifted by $0.2~\mathrm{\AA}^{-1}$ to the right with respect to the next lower pressure condition.}
\label{fig:dispersion_hugo}
\end{figure}

The adiabatic speed of sound can be computed from the thermodynamic relation
\begin{equation}
\label{eq:cs}
c_\mathrm{s} = \sqrt{\frac{\gamma}{\kappa_T \rho}},
\end{equation}
with the isothermal compressibilities $\kappa_T$ computed in Fig.~\ref{fig-statS4568}. The upper panel of Fig.~\ref{fig:dispersion_hugo} shows the linear trends computed from equation~\eqref{eq:cs}, as well as the disperison of the ion acoustic mode taken from the DSF shown in Fig.~\ref{fig:dynS_hugo_liquid}. It is apparent that the observed peak positions converge to the linear behavior for all conditions, while for the more extreme conditions, the hydrodynamic regime is reached at smaller $k$-values. The bottom panel of Fig.~\ref{fig:dynS_hugo_liquid} shows the peak position of the longitudinal current-current correlation spectrum across a wide range of wave numbers. The free-particle limit of a non-interacting system (see Eq.~\eqref{eq:freeparticle}) is given as a reference. The disperison above $10$~\AA$^{-1}$ for all conditions is well approximated by this limiting behavior.

\begin{figure}[!thb]
\center{\includegraphics[angle=0,width=1.0\linewidth]{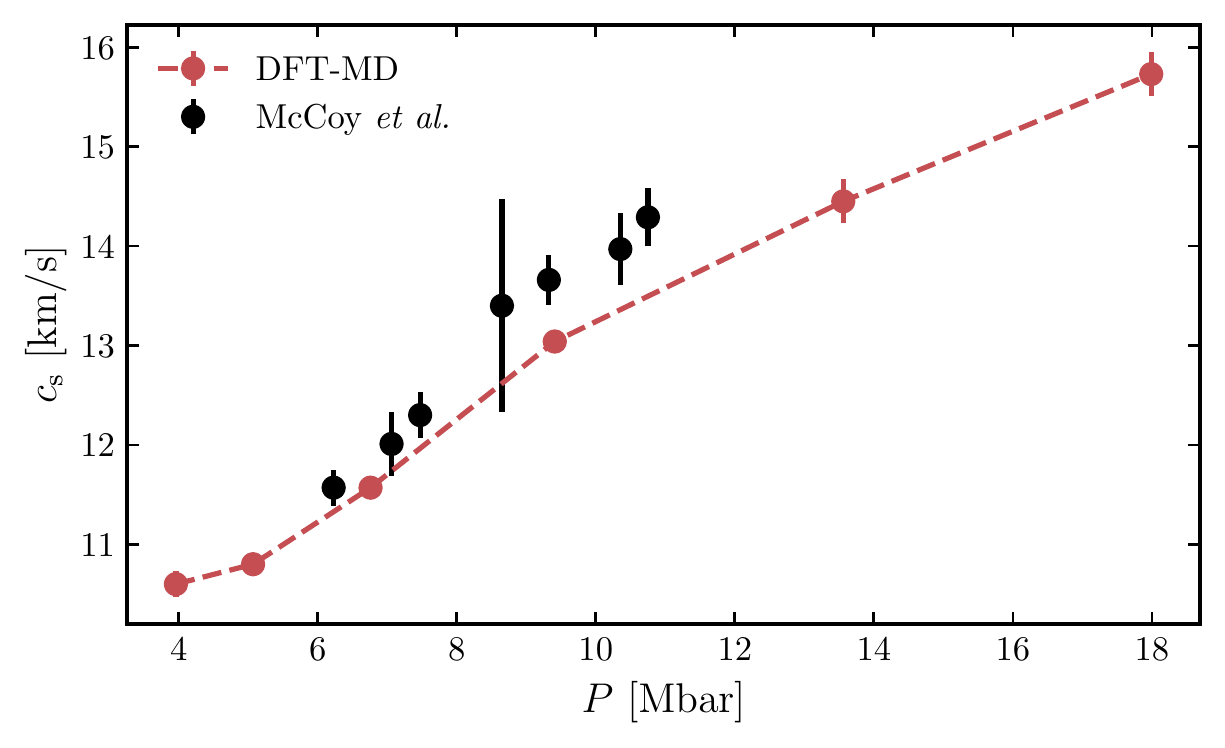}}
\caption{Adiabatic speed of sound computed from the thermodynamic relation~\eqref{eq:cs} compared to experimental VISAR measurements by McCoy~\textit{et~al.}~\cite{McCoy2017} for shock-compressed copper. 
}
\label{fig:sound_hugo}
\end{figure}

The adiabatic speed of sound is also accessible during shock wave experiments via VISAR measurements~\cite{Barker1972,Li2018}, where the time it takes the shock wave to travel through the target is recorded. 
First measurements of the speed of sound in copper for this pressure region were observed by McCoy \textit{et al.}~\cite{McCoy2017} during a shock compression experiment. The inferred pressure-density conditions, amongst others, are shown in Fig.~\ref{fig:hugo_exp} and agree within error bars with our simulations. In Fig.~\ref{fig:sound_hugo}, we show the experimentally determined adiabatic speed of sound compared to the values computed through Eq.~\eqref{eq:cs}. All of the simulation data points lie slightly below the experimentally observed results. As the principal Hugoniot becomes steeper in the $P-\rho$ plane, the adiabatic speed of sound appears to flatten out towards higher pressures. Unfortunately, the experimental data does not extend to these pressure to verify this trend.

\section{Conclusion}
\label{sec:concl}

In this work, we performed an extensive analysis of shock-compressed copper using DFT-MD simulations and MD simulations driven by high-dimensional neural network potentials. 
By analyzing the ion-ion structure factor, we showed that our DFT-MD simulations are able to accurately describe the phonon spectrum of solid copper. Likewise, our analysis of the dynamic electrical conductivity in terms of the Kubo-Greenwood formula and linear response TD-DFT yielded close agreement with existing experimental data. Furthermore, we computed the static and dynamic ion-ion structure factor of liquid copper near the melting line. The agreement with diffraction data was observed to be excellent and the width of the thermal mode agreed well with experiments at wave numbers around the first correlation peak.

The Hugoniot curve was computed from several isotherms upto $60~000$~K and 18~Mbar and compared to predictions by the SESAME EOS. Good agreement in the pressure-density plane was achieved between DFT-MD, SESAME and experiments upto 4~Mbar. Differences in the temperature between DFT-MD and SESAME along the Hugoniot were identified, and the resulting shift of the melting point to higher pressures was highlighted. We observed the hardening of phonon spectra in the solid regime of the Hugoniot compression and, analogously, studied the shift of ion acoustic modes to higher excitation energies. Phonon hardening is currently lively debated and recent work has shown that phonons can be resolved at free electron laser facilities~\cite{McBride2018,Descamps2020,Wollenweber2021,Descamps2022}, enabling future direct observations of phonon hardening in shock compression experiments.
We found the adiabatic speed of sound along the Hugoniot to be slightly underestimated by DFT-MD relative to recent experimental results.
We provided \textit{ab initio} predictions for the evolution of phonon and ion acoustic modes during shock compression of copper, as well as adiabtic speeds of sound for pressures beyond those previously reached by McCoy \textit{et~al.}~\cite{McCoy2017} We hope this inspires further high-pressure shock compression studies, coupled with high resolution x-ray scattering to resolve the ion dynamics of copper under these conditions.

\begin{acknowledgments}
We thank Emma McBride, Luke Fletcher and Siegfried Glenzer for helpful discussions. The DFT-MD and NN-MD simulations and further analysis were performed at the North-German Supercomputing Alliance (HLRN) and the ITMZ of the University of Rostock. MS and RR thank the Deutsche Forschungsgemeinschaft (DFG) for support within the Research Unit FOR~2440. This work was partially supported by the Center of Advanced Systems Understanding (CASUS) which is financed by Germany’s Federal Ministry of Education and Research (BMBF) and by the Saxon State Government out of the state budget approved by the Saxon State Parliament.
\end{acknowledgments}

\appendix
\section{Validation sets of NN potentials}

\begin{figure}[!thb]
\center{\includegraphics[angle=0,width=1.0\linewidth]{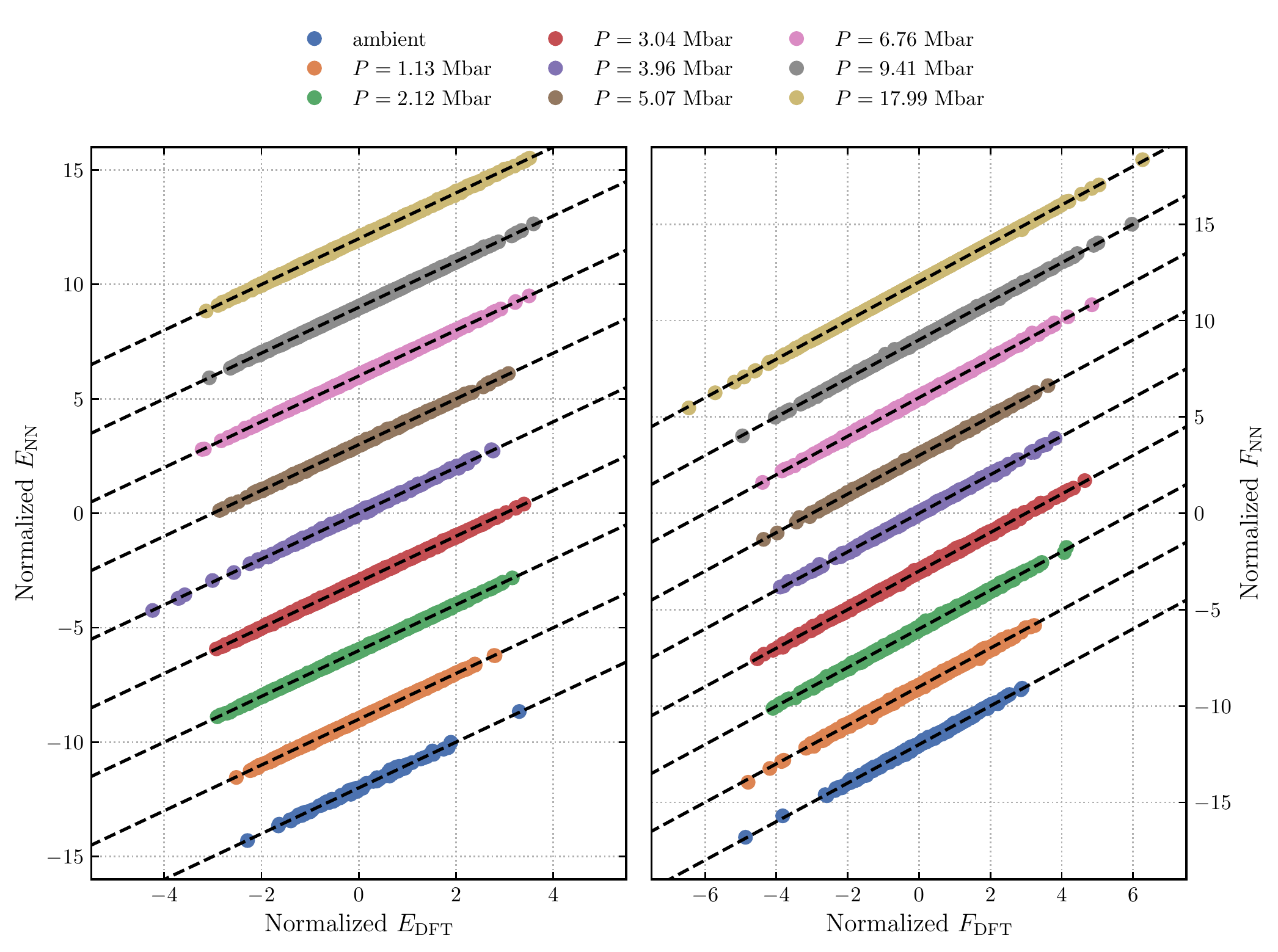}}
\caption{Validation sets of the NN potentials for energies and forces computed by DFT. As a reference, the black dashed lines show perfect correspondence between the NN potential and DFT.}
\label{fig:val_sets}
\end{figure}

We show the validation sets of forces and energies computed via DFT for all the conditions treated in this work in Fig~\ref{fig:val_sets}. Perfect predictions by the NN potential, corresponding to the straight lines $E_\mathrm{NN} = E_\mathrm{DFT}$ and $F_\mathrm{NN} = F_\mathrm{DFT}$, are indicated by the black dashed lines. Here, $E_\mathrm{DFT}$ and $F_\mathrm{DFT}$ are the energies and forces computed through DFT, where $F_\mathrm{DFT}$ can be any cartesian component of the force vector. Analogously, $E_\mathrm{NN}$ and $F_\mathrm{NN}$ are the forces predicted by the NN potential.

\newpage

\bibliography{copper_main.bib}

\end{document}